\documentclass[onecolumn,linenumbers,trackchanges]{aastex6}
\usepackage{times}
\usepackage{graphicx}
\usepackage{subfigure}
\usepackage{txfonts}
\usepackage{color}
\usepackage{booktabs}
\usepackage{multirow}
\usepackage{setspace}
\usepackage[flushleft]{threeparttable}
\usepackage{booktabs}
\usepackage{tablefootnote}
\usepackage{listings}

\usepackage{natbib}
\newcommand*\kms{\hbox{km~s$^{-1}$}}

\begin{document} 

\title{Undercover EUV solar jets observed by the Interface Region Imaging Spectrograph} 
\author{N.-H. Chen$^{1\ast,2}$ and D. E. Innes$^{2}$}
\affil{ $^{1}$Korea Astronomy and Space Science Institute, Daejeon, South Korea\\
$^{2}$Max-Planck-Institut f\"ur Sonnensystemforschung, D-37077 G\"ottingen, Germany\\}

\begin{abstract}

It is well-known that extreme ultraviolet (EUV) emission emitted at the solar surface is absorbed by overlying cool plasma. Especially in active regions dark lanes in EUV images suggest that much of the surface activity is obscured. Simultaneous observations from the Interface Region Imaging Spectrograph (IRIS), consisting of UV spectra and slit-jaw images (SJI) give vital information with sub-arcsecond spatial resolution on the dynamics of jets not seen in EUV images. We studied a series of small jets from recently formed bipole pairs beside the trailing spot of active region 11991, which occurred on 2014 March 5 from 15:02:21 UT to 17:04:07 UT. There were collimated outflows with bright roots in the SJI 1400 {\AA} (transition region) and 2796 {\AA} (upper chromosphere) that were mostly not seen in AIA 304 {\AA} (transition region) and AIA 171~\AA\ (lower corona) images. The Si IV spectra show strong blue-wing but no red-wing enhancements in the line profiles of the ejecta for all recurrent jets indicating  outward flows without twists. We see two types of Mg II line profiles produced by the jets spires: reversed and non-reversed. Mg II lines remain optically thick but turn into optically thin in the highly Dopper shifted wings.The energy flux contained in each recurrent jet is estimated using a velocity differential emission measure technique which measures the emitting power of the plasma as a function of line-of-sight velocity. We found that all the recurrent jets release similar energy (10$^8$ erg cm$^{-2}$ s$^{-1}$ ) toward the corona and the downward component is less than 3\%.
\end{abstract}


\section{Introduction}
Jet-like eruptions, occur in coronal holes, active regions, and the quiet Sun over a broad range of spatial and temporal scales \citep{1992shibata,1996Shimojo,2007cirtain,2010Moore,2016innes}. They have been observed at wavelengths ranging from hard X-ray (HXR) to white light, revealing their multi-thermal plasma constitution \citep{2013chen}. An inversed-Y or Eiffel-tower structure characterizes their standard morphology and they are typically seen near cancelling  and newly emerging magnetic fields \citep{1996Shimojo, 1996canfield, 2015cheung}. Therefore they are thought to be initiated by magnetic reconnection \citep{1992shibata, 2007shibata}.     

Jets are often associated with cool plasma surges \citep{1996canfield,1999chae, 2007jiang, 2008chifor, 2015sterling, 2016mulay}. \citet{1996canfield} observed  blue-shifted H$\alpha$ surges adjacent to X-ray jets, associated with moving magnetic features around a sunspot. \citet{1999chae} found recurring extreme ultraviolet (EUV) jets with correlated  H$\alpha$ surges near cancelling magnetic flux in a decaying active region. Recently, \citet{2016zeng} reported co-spatial H$\alpha$, EUV and HXR footpoint emission and well-correlated H$\alpha$ and UV spires in small-scale chromospheric jets observed by the New Solar Telescope (NST). In these coordinated observations, X-ray/EUV jets and H$\alpha$ surges are seen from regions with evolving magnetic fields and are characterized by common features, such as a point-like brightening at the footpoints and a nearly collimated spire. These studies have shown the close association between X-ray/EUV jets and H$\alpha$ surges, although a few observational discrepancies were also noted, such as a time delay between the H$\alpha$ surges and EUV jets \citep{2007jiang}. 
Other cool eruptions, such as Ca II jets, are notable particularly because they share a similar inversed Y-shaped scheme as the hot X-ray jets in the corona but are much smaller and occur in the chromosphere \citep{2008nishizuka}. Furthermore, most of the UV/EUV jets have been seen  at the edge of H$\alpha$ surges \citep{1996canfield, 1999chae}.  This could be due to inhomogenous density or temperature distributions along the outflows or that the cool and hot plasma were ejected along adjacent field lines \citep{1995yokoyama, 2005anzer}. The acceleration process may be similar both in cool and hot eruptions, but the detailed processes have not been clearly determined. 

To investigate the nature of the jet-like eruptions, we use data from the two solar missions: SDO \citep{pesnell2015solar} and IRIS \citep{2014IRIS} which provide images, spectra and magnetic field strengths from features in the photosphere, chromosphere, transition region and corona. The Atmospheric Imaging Assenbly \citep[AIA;][]{2012lemen} on SDO consists of seven EUV and two UV channels with a 12 s and 24 s cadence and $1\arcsec.5$ spatial resolution of the entire solar disk. The lines are centered at 94, 131, 171, 193, 211, 304, 335, 1600, and 1700~\AA. We use images from the Heliospheric and and Magnetic Imager \citep[HMI;][]{2012schou} on SDO  to investigate the underlying photospheric line-of-sight (LOS) magnetic field and intensity. The IRIS spectrograph observes in two bands: FUV (1332-1406 {\AA}) and NUV (2783-2834 {\AA}) with a spatial and spectral resolution of 0\arcsec.33 and 26 m{\AA}. The slit-jaw images (SJI) can be obtained simultaneously with spectra in the wavelength regions 1330, 1400, 2796, and  2830 {\AA} with a spatial resolution of 0.16\arcsec\ pixel$^{-1}$.
and a cadence of a few seconds. 

The aim of this paper is to describe the initiation and evolution of recurrent jets from a region with moving magnetic features just outside the sunspot penumbra.
In the following, we present a multi-wavelength analysis of the recurrent jets by using  images from IRIS and AIA and spectra from IRIS. In section 2, the detailed spatial evolution is investigated from the simultaneous imaging of IRIS and AIA. The corresponding line profiles and the energy of the recurrent jets are described in section 3. Finally, the summary and discussion are given in section 4. 

\section{Observations}
\subsection{Overview of observations} 
\begin{table}[h]
\centering
\caption{IRIS observations in OBS 3860260418.}
\resizebox{0.85\columnwidth} {!} {
\begin{tabular}{ll}
\hline
Duration (UT)              & 14:49:52 --- 17:45:14                                        \\ \hline
\# Raster                 & 165                                                        \\
Raster Cad. (s)            & 64                                                         \\
\# Exp./ Raster                   & 4                                                          \\
Exp. Time				& 15.0  						\\
Spectral lines (\AA)       & C II 1336, Fe XII 1349, O I 1356, Si IV 1403, Mg II k 2796 \\
SJIs (\AA)/ cadence (s) / exposure (s) & 1400/39/15 ; 2796/32/15 ; 2832/191/15                               \\ \hline
\end{tabular}}
\end{table}

The events under study were a series of jets that recurred near the periphery of the trailing sunspot of AR11991 when it was near the disk center on 2014 March 5. The IRIS slit crossed the jets from east to west in a large, sparse, 4-step raster covering a field-of-view (FOV) $3\arcsec\times119\arcsec$ in 64 s with $1\arcsec$ step size and exposure time 15 s (Table 1). We concentrate on the IRIS Si IV spectra, IRIS SJI 1400 and  2796 images, and AIA 171, and 304 {\AA} (hereafter, A171 and A304)  observations with the objective of investigating the behaviour of each recurrent jet. The SJI 1400~{\AA} images are dominated by Si IV line emission from the transition region and continuum emission from the low chromosphere. The 2796 {\AA} images, dominated by Mg II k lines, are used to study the corresponding upper chromospheric behaviour of the jets.  As for the two channels of AIA: A171 is emitted primarily from plasma with a formation temperature at $6\times10^{5}$ K so should be good for tracking the coronal evolution of the jets and A304, formed around the same temperature as the Si IV lines, is dominated by He II emission. Absorption by overlying neutral plasma may cause a darkening of the EUV emission from the jet and footpoint.  To coalign these images, first, we scaled the AIA and HMI images to match the FOV and pixel size of the IRIS SJI images. Second, we coaligned the 2832 {\AA} and HMI continuum images on prominent features, such as the sunspot. We further matched the 1600 {\AA} with the HMI continuum images, and also the 2832 {\AA} with the other SJI bandpass images separately. Then the SJI images can be coaligned to the rest of co-temporal AIA images. We also examined the LOS  HMI magnetograms for  clues on the initiation of the jets.

Here we use the strong Si IV lines and also Mg II lines for the spectral study and present the line profiles in the jet and loop. The shape of the Si IV line profile in the jet's spire often consists of two components: a highly blue-shifted component, namely the moving outflow, and a slight red-shifted component, representing the background bulk plasma. To begin with the  characteristics of the line profiles, such as the line width are obtained from  double-Gaussian fits. More results will be discussed in section 3.

\subsection{Recurrent jets}

\begin{figure*}[htp] 
\begin{center}
 \includegraphics[width=.86\textwidth,clip=true, trim = .9cm 0.5cm 1cm 4.75cm]{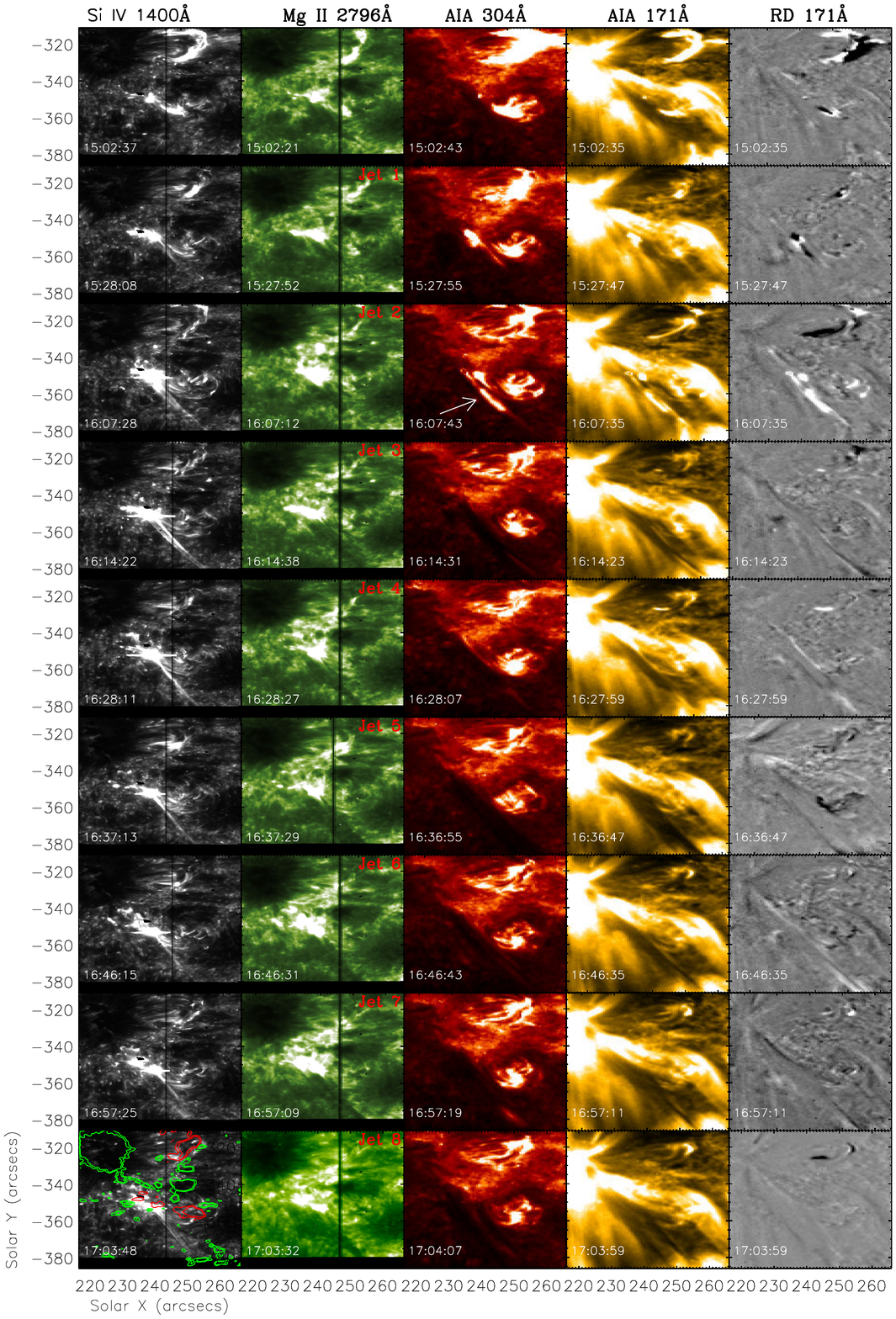}
 \caption{Context images of recurrent jets in AR 11991 on 2014 March 5. Panels from left to right show coaligned IRIS 1400 and 2796 {\AA} slit-jaw images, AIA 304 and 171 {\AA} images and 171 {\AA} running-difference images.  The jets are labeled as Jet 1, 2 etc. and the associated magnetic bipoles are overplotted on the left-most frame in which the red/ green contours mark the positive/ negative polarity with the 10 \% and 30 \% of maximum counts ($\pm500$~G) in the HMI LOS magnetogram. The white arrow drawn on the image taken at time 16:07:43 UT points to an abnormal brightening (see text). }
\end{center}
\end{figure*}

\begin{figure*}[htp] 
\begin{center}
 \includegraphics[width=0.87\textwidth,clip=true, trim = 0.85cm 1cm 2cm 4cm]{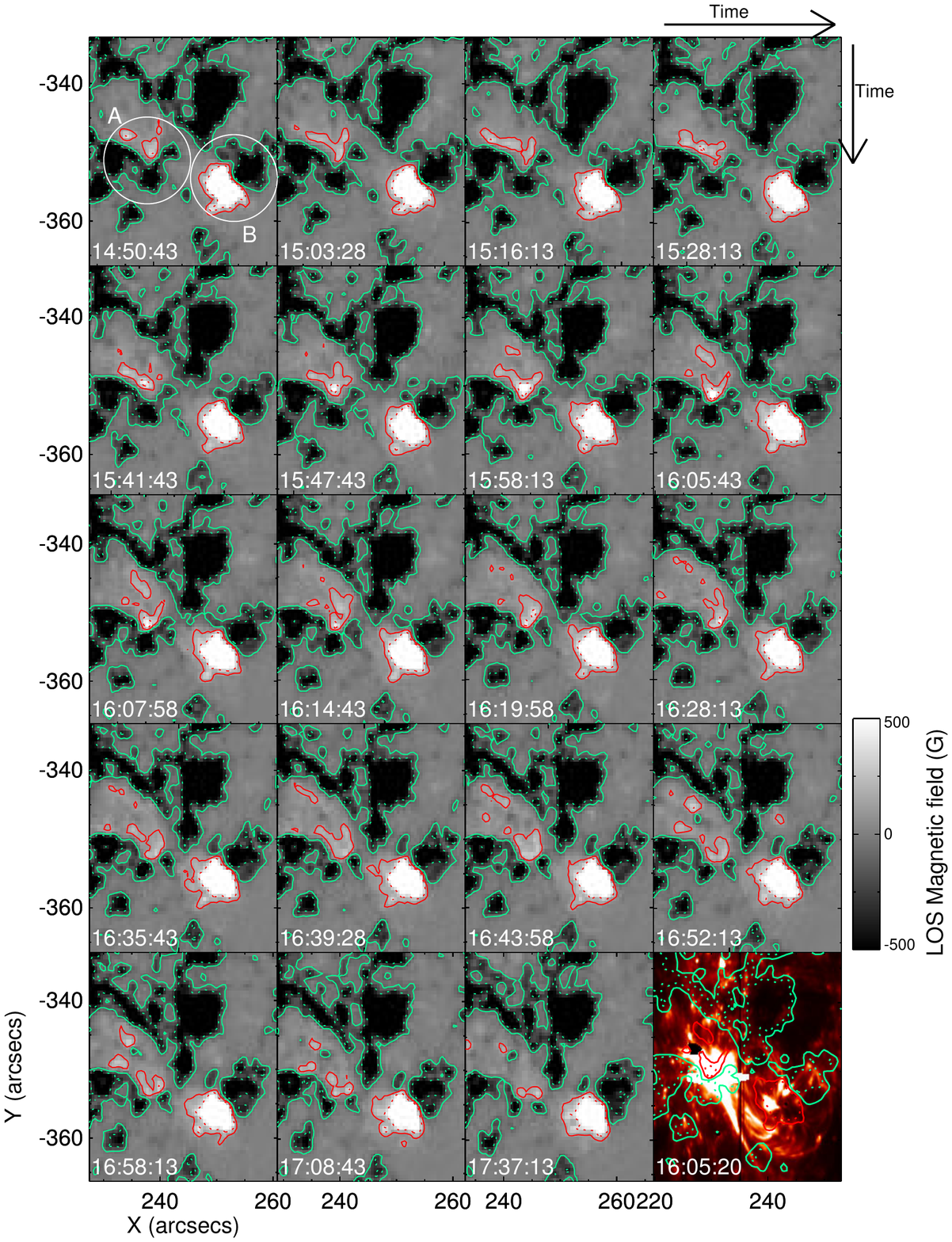}
 \caption{HMI LOS magnetograms during the IRIS observation (time starts from left to right, top to bottom as shown by the black arrows). The red/ green contours mark the positive (white)/ negative (black) polarity with the 10 \% (solid lines) and 30 \% (dash lines) of maximum counts ($\pm500$~G). The right-most bottom frame gives the position of 1400 {\AA} jets. Circle A surrounds identified biploes related to the jets and B surrounds the positive field loop footpoint.}
\end{center}
\end{figure*}

\begin{figure*}[htp] 
\begin{center}
  \includegraphics[width=.82\textwidth,clip=true, trim = .9cm 6cm 1cm 12cm]{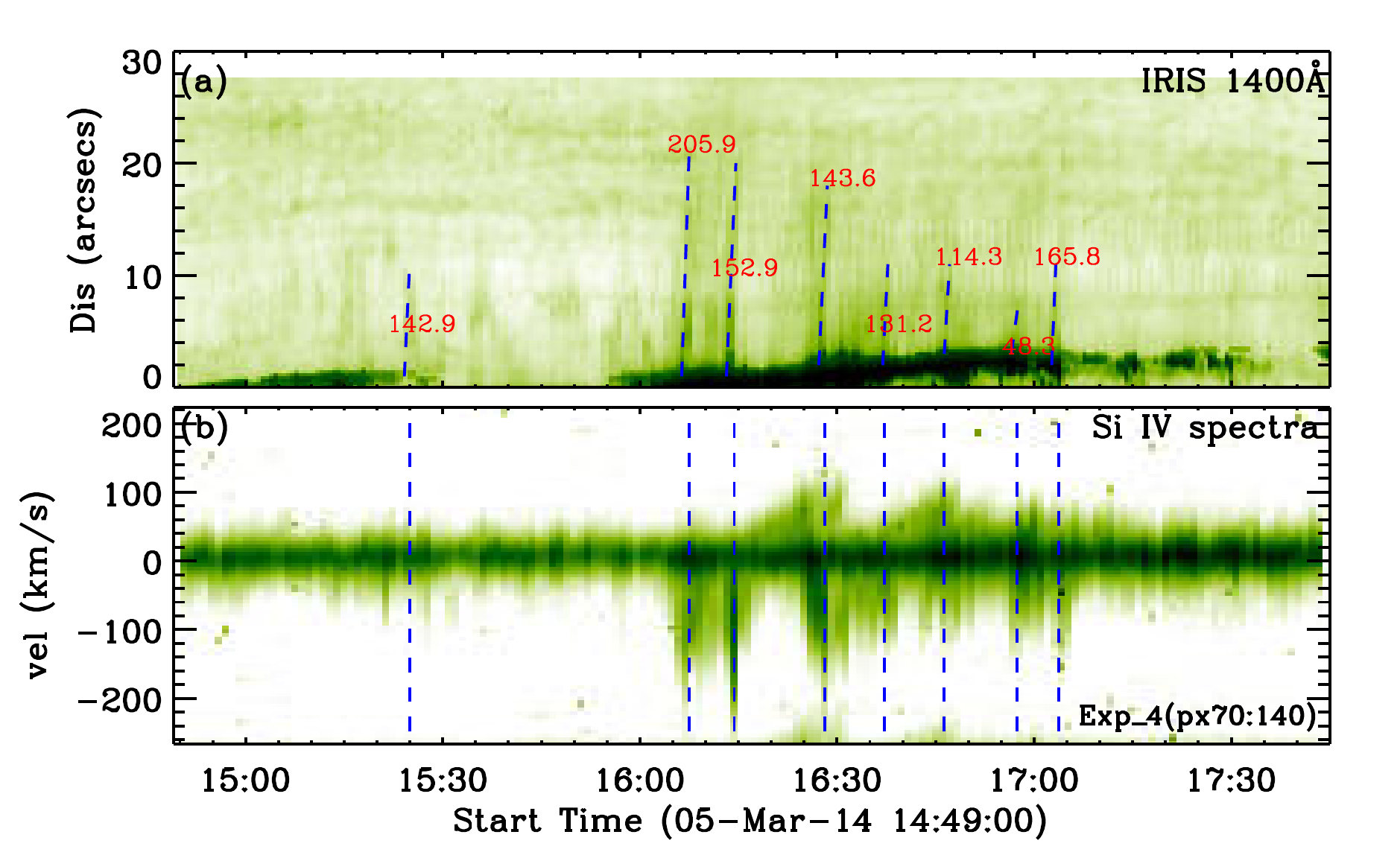}
  \includegraphics[width=.6\textwidth,clip=true, trim =0.5cm 0.5cm 0.5cm 11cm]{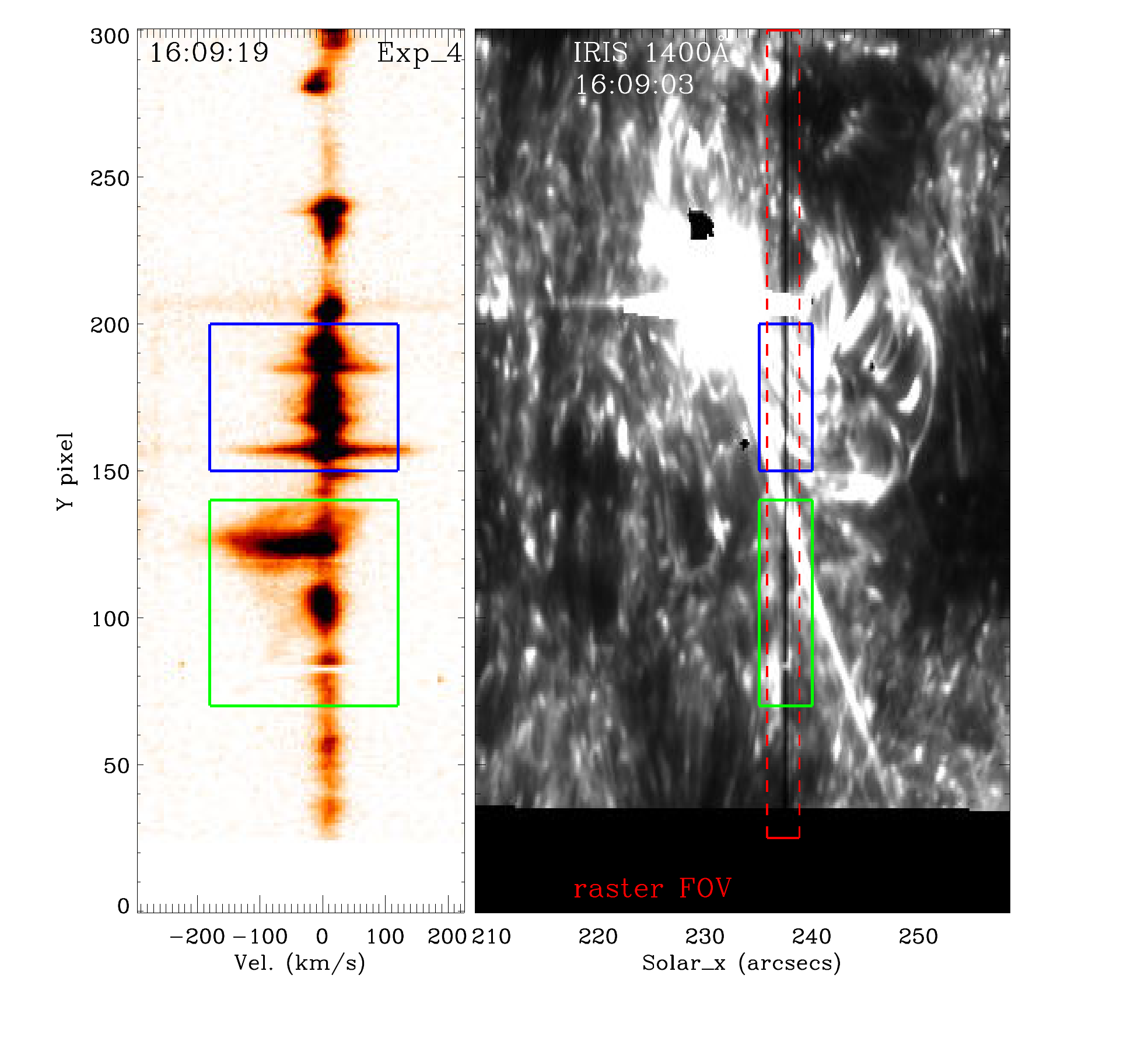}
  \caption{Top: (a) Space-time map of SJI 1400~\AA\ emission along the jet. (b) Temporal evolution of  Si IV 1402~\AA\  spectra. In the  SJI images, the jet is traced via a line-cut along the axis of the spire. Dash lines are the slope of each jet used to calculate the speed of jets and each is tagged with a speed in \kms. In the spectral images, the line intensity is averaged for 4th exposure over the region of the jet (green box in the bottom plots). The time of each jet is marked by the blue dash lines. Bottom: Regions of interest in spectra image. It illustrates our selection of  specific pixels for estimating of LOS jet speed. The green box in the spectral  image (left) is selected based on the positions of the spire shown on the slit-jaw images (right). The red dashed box shows the FOV covered by the slit in one four-step raster. The blue box covers the region of loop emission. Note these spectra typically have both high velocity red and blue wings. }
\end{center}
\end{figure*}

The development of the jets is best studied in the 1400 {\AA} images.  Figure 1 shows all the recurrent jets (labeled as Jet 1, 2 etc) from 15:02:21 UT to 17:04:07 UT in the 1400, 2796, 304, 171, and 171 {\AA} running-difference images. The LOS magnetogram contours are overlaid on the leftmost bottom frame to show the associated magnetic bipoles. Eight jets were observed coming from the same location during the  three hours observation. Jet 1 was the faintest. All of them show a clear collimated long spire with bright footpoints, namely a typical inversed-Y shaped geometry, which is in agreement with the standard solar jet model \citep{1996Shimojo}. Although the contrast of the 2796 {\AA} images is lower, the identical bright footpoints are  visible in these images as well. The spire itself in the 2796 {\AA} images is actually dimmer than the surroundings. They have the same length as the bright ones but are morphologically wider. The dissimilar appearance is due to the line formation process. While the strong emission of Si IV line is proportional to the column density of transition region plasma along the LOS,  the Mg II k line, emitted by cooler plasma, is optically thick and its intensity is not a direct reflection of the column of Mg II along the line-of-sight. The intensity is given by the source function at optical depth unity. Since the jets are darker than the surroundings, the source function and possibly the temperature of the Mg II plasma is cooler in the jet than in the surrounding. Investigation of the line profiles is required to understand the Mg II emission.  The reason that most the jets are not visible in the EUV images is probably because the jet EUV emission was absorbed by overlying cool plasma.

These observations demonstrate the importance of the UV observations for revealing  heated jet plasma below the overlying cold material and for identifying the jet footpoints. Before the IRIS-era, EUV together with H$\alpha$ images were often used to diagnose the evolution of solar jets and the accompanying surges \citep{2015chenJ, 2014zhang, 2014adams} so it is interesting to see how much additional information on the jet morphology is revealed by including the UV images. In the jets described here, typical features such as the bright footpoints and spire, are only partially visible in the EUV channels, which is also evidenced in the sequences of running-difference (RD) A171 images (column 5). The RD images are created from  two consecutive images taken 12 s apart and show the intensity enhancement/decrease of the observed A171 emission in the eruptions. Only two of the jets (Jet 1 \& 2) show simultaneous enhancements of the EUV with the  1400 {\AA} jets. The other jets show no obvious response (mostly grey) in the EUV, even at their footpoints. The most likely interpretation is surge plasma absorbing the EUV as indicated by the darkening in the 2796~\AA\ images.  It is noted that there is also an abnormal intensity enhancement (white arrow) in Jet 2 resulting from the superposition of an additional jet launched from a different starting point along the path of the other jets (see movie  aia\_2.mp4).

Figure 2 shows the temporal evolution of the photospheric LOS magnetic fields near the base of the recurrent jets. There were two obvious bipole regions labelled as A and B (white circles) on the south west of the trailing spot of AR 11991 (see bottom left image in Figure 1). The positive (negative) flux concentrations are outlined in the red (green) contours with the 10{\%} (solid), 30{\%} (dot) of the maximum magnetic field strength. In the region A, the positive flux  separated and migrated gradually in a west-south direction as they emerged. Their migration was terminated when they  cancelled  with the negative flux. The recurrent jets were launched from  region A which was also the site of the flux cancellation. Region B was more compact and slightly decomposed at the edge of the negative pole where the footpoint of the loops were situated. 

To estimate the speed of each jet, we make a distance-time map, shown in Figure 3(a).  The jets are traced via a line-cut along the axis of the spire in the SJI 1400~\AA\ and appear as  dark elongations (indicated by blue dashed lines) in the intensity-reversed map. The speed of each jet ranged from 50 to 200~\kms\ which is identical to the speed of the jets observed by TRACE in the EUV \citep{1999chae}. We also measure their LOS velocity in the  averaged 1402 {\AA} line intensity map shown in figure 3(b). The Si IV line intensity is averaged over the fourth exposure over the selected pixels (green box in the lower right hand panel) for the jet's spire. As each raster consists four 15 s exposures (plus CCD read-out time $\sim$1.5 s), the plot has a 64 s cadence. The Si IV 1402.773~{\AA} was used as the reference wavelength for the measurement of Doppler shifts.  The negative Doppler velocity gives the blue shift, namely the direction away from the Sun while the positive shows the red shift. Each jet instance was highly blueshifted ($>$100~\kms) except for the first jet which was missed by the slit (in the 4th exposure). The LOS velocity from spectra and the speed derived from the temporal evolution of SJI images are roughly equal, so the combination gives a jet speed between 70 and 300~\kms. The bottom frame of figure 3 shows an example of a SJI image and co-temporal 1400 {\AA} spectral image. 

\section{Time-dependent Spectroscopy}
\subsection{Spectra of jets}
Figure 4 gives an overview of the line profiles seen in Si IV and Mg II lines at selected times. To study the spectra during the jets, we selected the pixels where the slit crossed the spire and the loops connecting the footpoint of the jet to the nearby region B (see previous section). The cyan and purple lines in figure 4 show the selected pixels and the line profile taken simultaneously with SJI is shown on the right together with the spectral image. We use 2796.34 {\AA} and 2803.52 {\AA} as the rest wavelengths of the Mg II k and h lines respectively for measuring the Doppler shifts. The results of fitting the Si IV spectra are shown with a double-Gaussian at jet's spire after background emission from the continuum has been subtracted. In the Si IV spectra, every jet (blue solid line) had a highly blue-shifted asymmetric profile or component and almost no enhancement in the red wing, indicating that this is a collimated outflow without twists. The dominant blue component of each spire's profile extends to 200~\kms~and the width ranged from 23-48~\kms, suggesting a strong non-thermal upward outflow.  The width of their red wings is bounded between 3-40~\kms. During the jets, a single-Gaussian method is used to fit the narrow line profile at the corresponding loop pixels (red solid line). The loops' profiles are slightly red-shifted and some have intense peak emission which is about a factor two stronger than the spires' emission. 
 
To better understand the behavior of the Mg II lines during the jets, we consider the k to h intensity ratio in the line core and wings. The Mg II line formation was discussed extensively in \citet{2013a_leenaarts,leenaarts_2013b} with emphasis on the structure of the lines near the core but the profiles from the jet spire are far from typical due to flows along the jet (Fig. 4), so it is difficult to select line core and peaks as done by Leenaarts. Instead we have measured ratios of the k to h at specific Doppler shift in order to determine whether the emission is optically thin in the jet. As noted, the intensity in the jet core is significantly less than in the surroundings. This is because the jet is above the typical chromosphere and in the core, the line is optically thick so one only sees emission from the optical depth equals one layer (e.g. \citet{2013a_leenaarts}). In the wings, the intensities in the jet are higher than in the surroundings and it is hard to tell if the Mg II lines from the jet is optically thin or thick. If it is optically thin then the whole column in the jet would contribute emission on top of the background which is coming from the temperature minimum. If it is optically thick then the emission is coming from the optical depth equal one layer in the jet and the emission from the temperature minimum is not seen.

We therefore plot the ratio of the two lines at 0, $\pm50$ and $\pm80$~\kms~in the jet and at 0~\kms~in the neighboring loops. Away from the line core we plot both with and without background subtraction. The background values are taken as the intensity 0, $\pm50$ and $\pm80$~\kms~before jetting. We take the average value of intensity at the different velocities (both k and h lines) in three selected rasters ($\sim$3 mins) and the standard deviation is used to determine the minimum/maximum background values used in the line ratio computations. The ratio is very sensitive to the height of the lines above the background when the difference is small. Therefore  the accuracy of the ratio for weak emission is much less than for stronger emission.The error bars on the ratios reflect the difference between ratios computed with a maximum and minimum background level. The error bar is only applicable and shown when the line intensity is higher than the background level. In the core (0~\kms) of the jet's spire, the k to h ratio is between 1.2 to 1.5 and reduces slightly (1.1$\sim$1.28) at loop pixels (see figure 5a). In the wings of the jets, the ratio with background ranged from 1.19 to 1.8 at $\pm50$~\kms and from 0.7 to 1.5 at $\pm80$~\kms. After the background subtraction, it is bounded by same range at -50~\kms~but drops down to less than 1 at ~+50~\kms. In the high Doppler velocity jet ($\pm80$~\kms), several rise up to 2, indicating that the spires are possibly optically thin. The relation of  the k and h lines intensity (figure 5d) reveals that Mg II lines at the core and wings of the jet are mostly optically thick even though the central reversal feature sometimes vanishes (e.g. Jet 6). But it could turn into optically thin in the high Doppler shifted wings.

The Mg II spectra at the loops show the standard line profiles with peaks on either side of the line core.  Most cases  have  a stronger blueward peak (two nearly equivalent peaks at jet 6 and 8) for both the h and k lines, signifying downflowing material above the optical depth equals one height \citep{leenaarts_2013b}. The peak separation is between 25 and 50~\kms~which is larger than the separation width of Mg II peaks in a plage region reported by \citet{2015carlsson}.

\begin{figure*}[htp] 
\begin{center}
 \includegraphics[width=.85\textwidth,clip=true, trim =0.2cm 1.6cm 0cm 0.3cm]{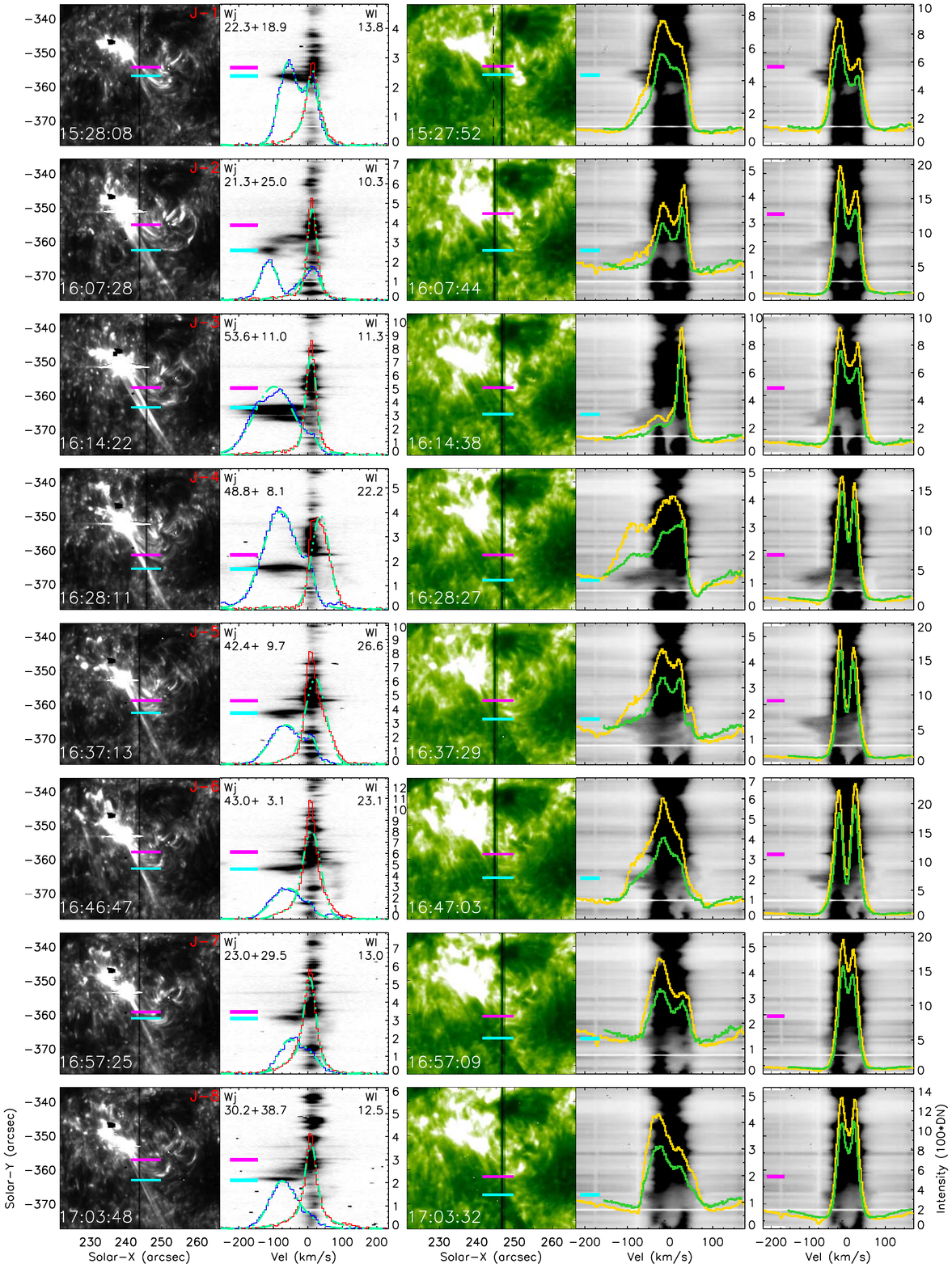}
 \caption{The Si IV (left) and Mg II (right) lines profiles. The spectral images of Si IV (around 1402 {\AA}) and Mg II  (2793-2806 {\AA}) at selected time are overplotted with the line profiles from the pixel that is marked with the a cyan (jet) or a purple (loop) bar on the left. The count of each line profile is shown on the right y-axis with units of DN. The co-temporal SJI images of Si IV and Mg II are shown on the left of the spectra with the locations of the selected profiles likewise indicated with the cyan and purple bars. Si IV (blue, red), Mg II k (yellow) and h (green) line profiles as a function of velocity shifted away from the line center (eg. 1402.77{\AA}, 2796.34 {\AA} and 2803.52{\AA} respectively). The light green dashed-dotted line is the best-fitting in double- (jet) and single-Gaussian (loop) methods and the fitting parameters of line width are given on the top of each subfigure. }
\end{center}
\end{figure*}

\begin{figure*}[htp] 
\begin{center}
 \includegraphics[width=.9\textwidth,clip=true, trim =1.2cm 9.5cm 0cm 1.5cm]{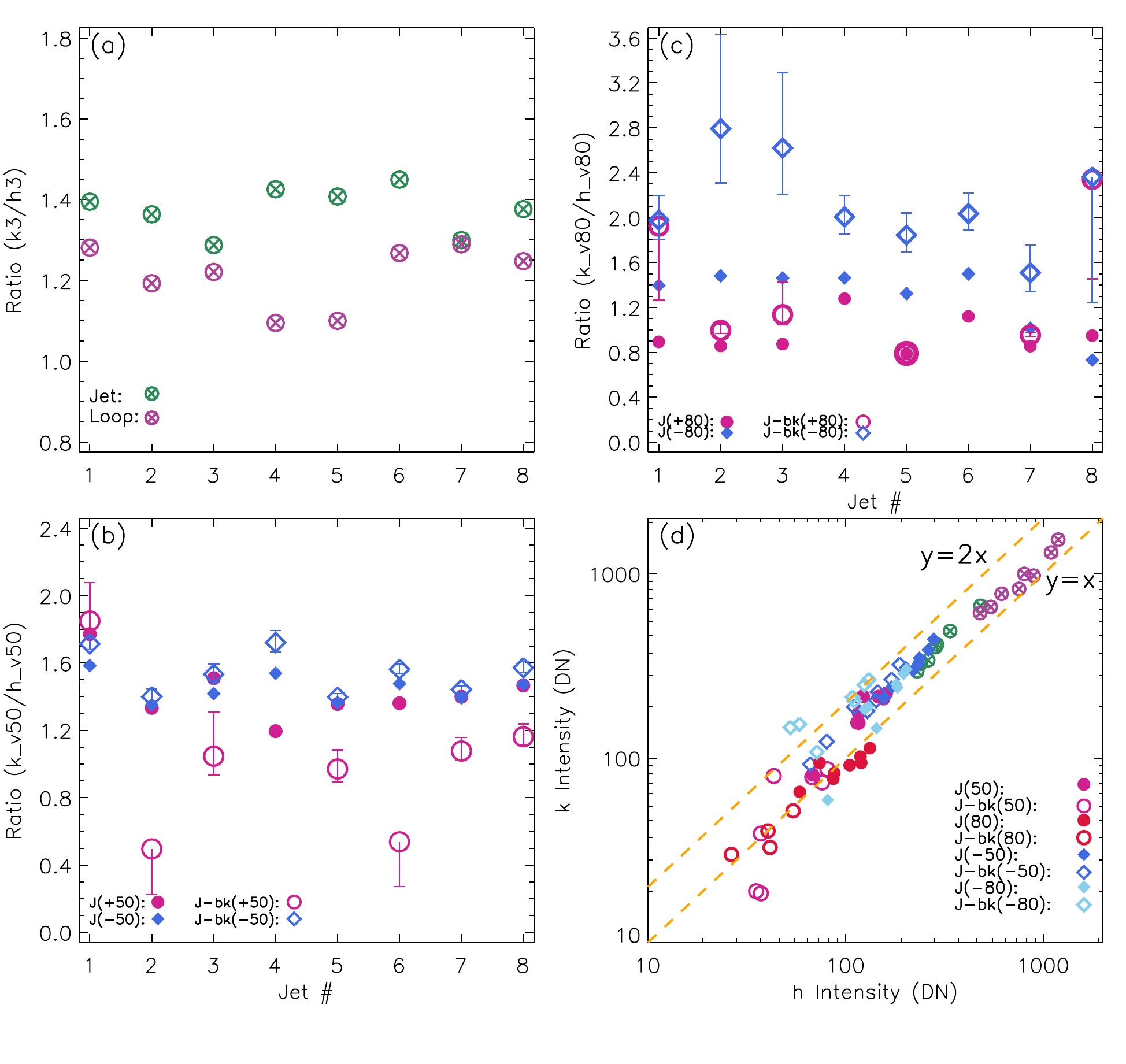}
 \caption{Mg II k/h line intensity ratios at (a)line center (vel.= 0~\kms), (b)vel.=~$\pm50$~\kms and (c)vel.=~$\pm80$~\kms. (d)The scatter plot of k line intensity vs. h line intensity at vel.=~0,~$\pm50$,~$\pm80$~\kms~with the orange dash lines, ratio of 1 and 2 shown .The green and purple circles with x are the ratio at the jet and neighboring loops respectively. The filled and open symbols (circle and diamond) in (b), (c) and (d) illustrate the ratio with and without background subtraction seperately. Crimson and violet red give the positive (redward) Doppler velocity. Deep and light blue show the negative (blueward) Doppler velocity. The background are taken as the intensity value of the vel. at 0, $\pm50$ and $\pm80$~\kms~before the jet occurred. The error bars in (b) and (c) are the minimum/ maximum ratio values due to the background subtraction. 
}
\end{center}
\end{figure*}

\subsection{Energetics of the recurrent jets}
In this section, we address the energy possibly contained in these recurrent jets. The analysis method adopted here is the Velocity Differential Emission Measure (VDEM) introduced by \citet{1995newton} for  flare events and \citet{1999winebarger, 2002winebarger} for explosive events in SOHO/SUMER observations. The VDEM gives a measure of the energy of the emitting plasma moving at the LOS velocity in the observed profiles (for more information, see \citet{1995newton} \& \citet{1999winebarger}). It is defined by
\begin{equation}
 \hbox{VDEM}=n^{2}_{e} G(T_e) \frac{ds}{dv} 
\end{equation}
where \textsl{$n_{e}$} is the electron density, $G(T_e)$ is the temperature, $T_e$, dependent emissivity function of the line, \textsl{s} is the distance along the LOS and \textsl{v} is the velocity. When deriving the VDEM from the observed spectrum, we assume that the plasma is moving along the LOS direction with uniform density and pressure during the exposure. Also the observed emission is assumed to have been emitted from the plasma at the temperature of the peak in the emissivity function. Since the propagation direction of the jets here is oblique and not only radially outward, the true energy flux would be larger. Before applying the method to IRIS observations, we  increased the photon counts (the minimum required photon counts  suggested in \citet{2002winebarger} is $>$ 1000 counts) by first summing the line profiles of three consecutive spatial pixels centered at the position with the maximum width along the slit (mostly around the jet's spire). We also took the thermal broadening of Si IV lines (6.88~\kms\ at $T_e = 8\times10^4$ K), non-thermal broadening (decided in each profile) and the instrumental broadening of IRIS (3.9~\kms , suggested in \citet{2014IRIS}) into account. The corresponding VDEM function is then produced after  deconvolution of the observed spectrum from which the moments of the velocity can be calculated. Next, we are able to estimate  the kinetic energy flux in the plasma outflow/infow by using equation (1) in \citet{2002winebarger}. 

Table 2 summarizes the results of the estimated energy flux of each jet. Three directional components have been determined, including the energy flux, kinetic energy flux and skewed energy flux. The directional energy flux is the total energy carried by the (in-/out-) flows which consists of the total thermal enthalpy and the total non-thermal energy flux in the certain direction. The kinetic energy component represents the energy contained in the flows moving with an average velocity while the skewed component indicates the high energy contribution, e.g. the high velocity term of the spectral line profiles (two ends of red and blue wings). Our calculations reveal that the average energy flux contributed toward the corona (upward) was $2\times $10$^8$ ergs~cm$^{-2}$~s$^{-1}$ and toward the chromosphere (downward) was $2.2\times $10$^6$ ergs~cm$^{-2}$~s$^{-1}$. Among the average upward energy flux, 39.6\% was from the kinetic energy flux. The downward kinetic energy was 1\% of the average downward energy flux. It is noted that the kinetic is $<$ 50\% of the total upward energy in each jet, suggesting that more than half of energy goes to thermal energy that heated the surrounding transition region plasma. The energy in the downward direction is very small.

\begin{table}[h]
\caption{Directional energy flux}

\begin{threeparttable}
\resizebox{1\columnwidth} {!} {
\begin{tabular} {|r|c|c|c|c|c|c|c|c|} \hline
\multicolumn{1}{|c|}{\begin{tabular}[c]{@{}c@{}}Energy flux\\  ($10^{_{6}}$ ergs~cm$^{-2}$~s$^{-1}$)\end{tabular}} & J1                                            
& J2     & J3       & J4        & J5      & J6        & J7         & J8           \\ \hline
 $E_{up}$    & 48.02    & 229.93   & 469.93      & 348.04     & 138.67     & 123.92  & 57.01   & 176.53    \\ \hline
 $E_{down}$      & 1.80    & 1.98    & 0.16  & 2.96 & 1.01   & 6.92  & 2.20   & 0.39        \\ \hline
 $Ek_{up}$    & \begin{tabular}[c]{@{}c@{}}11.23\\ (23.3\%)\end{tabular} & \begin{tabular}[c]{@{}c@{}}67.77\\ (29.4\%)\end{tabular} & \begin{tabular}[c]{@{}c@{}}218.59 \\ (46.5\%)\end{tabular} & \begin{tabular}[c]{@{}c@{}}141.93\\ (40.7\%)\end{tabular} & \begin{tabular}[c]{@{}c@{}}48.75\\ (35.1\%)\end{tabular} & \begin{tabular}[c]{@{}c@{}}42.68\\ (34.3\%)\end{tabular} & \begin{tabular}[c]{@{}c@{}}16.70\\ (29.2\%)\end{tabular} & \begin{tabular}[c]{@{}c@{}}83.43\\ (47.2\%)\end{tabular} \\ \hline
 $Ek_{down}$($10^{_{4}}$)                                                                                      & \begin{tabular}[c]{@{}c@{}}4.31\\ (2.4\%)\end{tabular}   & \begin{tabular}[c]{@{}c@{}}4.60\\ (2.3\%)\end{tabular}   & \begin{tabular}[c]{@{}c@{}}0.017\\ (0.11\%)\end{tabular}   & \begin{tabular}[c]{@{}c@{}}0.42\\ (0.14\%)\end{tabular}   & \begin{tabular}[c]{@{}c@{}}0.44\\ (0.44\%)\end{tabular}  & \begin{tabular}[c]{@{}c@{}}5.60\\ (0.81\%)\end{tabular}  & \begin{tabular}[c]{@{}c@{}}2.47\\ (1.1\%)\end{tabular}   & \begin{tabular}[c]{@{}c@{}}0.068\\ (0.2\%)\end{tabular}  \\ \hline
\end{tabular} }
\begin{tablenotes}
   \item 1: $E_{up}$, $E_{down}$, $Ek_{up}$ and $Ek_{down}$ are the upward/downward total energy and upward/downward kinetic enegy separately. 
    \item 2: The value of each energy flux is the multiples of 10$^6$, except for the $Ek_{down}$ that is for 10$^4$. 
    \item 3: The precentage of $Ek_{up}$ to $E_{up}$ is gaven underneath and so does the downward direction.
   \end{tablenotes}
\end{threeparttable} 
\end{table}

\section{Discussion and summary}
We studied the spatial and temporal evolution of recurrent jets both with AIA and IRIS observations in detail. These jets recurred above the active flux cancellation regions in the periphery of AR 11991. The spectroscopic studies of these jets suggest collimated outflows without  twists. Although the formation of Si IV and Mg II lines are not the same, the regular features of a standard jet were commonly visible in the slit-jaw images. We also found  discrepancies in the morphologies  of the jets in the  UV transition region and EUV transition region and corona emission (as shown in figure 1). One explanation is that pre-existing cool material lay between the jets and observer which caused the absorption of the EUV at short wavelengths \citep{2003innes, 2005anzer, 2009pontieu}. Another possibility is that the cool jets/ material were accelerated and ejected because of the slingshot effect caused by the magnetic reconnection if the cool plasma was situated around the reconnection sites \citep{1995yokoyama, 2008nishizuka}. The coexistence of cool and hot plasma ejections could also be due to the cooling of previous hot ejections/ material suggested in previous works \citep{1994schmieder,1999alex,2007jiang}. For our case here, the dim coronal emission appeared cospatially and cotemporally in the observed wavelengths which makes the cooling mechanism unlikely. And most of the cool/hot plasma around the reconnection sites would be exhausted after first ejection (drop by factor of 2 in the following ejection \citep{2010archontis}) so it could be difficult to supply enough cool plasma in the following ejections within the repetition period (5-15 mins) of jets 2-8 here. It is worth noting that dim coronal ejections recurred at the same region of AR 11991 throughout the whole day. The driving mechanism of such long duration dark/dim ejections still needs further investigations.

The line formation of Mg II lines involves complicated physics which is greatly influenced by the inhomogeneous solar stratification. The appearance of the central reversal behaviour in Mg II lines might be related to the ratio between collision to radiative de-excitation, the density and the temperature if a simplified radiative transfer model is assumed \citep{2013a_leenaarts}. We found some profiles in spire position were non-reversed with more symmetric profiles, e.g. the core in emission, but some were reversed. It might relate to the  enhancement of the density or temperature in each jet-associated reconnection process. Some flare observations \citep{2015liu} suggest that the non-reversed profiles can be due to high coronal pressures that could be attributed to the evaporation flows of jets and the reversed ones is often associated with strong heating. We found the correlation of k to h line intensity ratio at the line core and wings remains optically thick, however, it might become optically thin under the highly Dopper shifted condition. 

 As noted in Si IV lines spectra (figure 4), the broader blueward component is, the stronger energy flux would be. Energetics of jets derived in Si IV lines demonstrated a similar magnitude in the upward energy flux with a downward energy flux of less than 3\%. Our case show the average energy flux of these jets is two-order-magnitude larger than the explosive events reported in SUMER observations. The energy of those transition region jets or small explosive events, however, is not the major source to heat the corona or chromosphere. It is still significantly evidenced that these undercover solar jets, contained almost similar energy (or mass) as other observed jets, might be occulted frequently in current EUV observations.

\bibliographystyle{aasjournal} 
 \bibliography {jet_draft}

\begin{thebibliography}{}
\expandafter\ifx\csname natexlab\endcsname\relax\def\natexlab#1{#1}\fi

\bibitem[{{Adams} {et~al.}(2014){Adams}, {Sterling}, {Moore}, \&
  {Gary}}]{2014adams}
{Adams}, M., {Sterling}, A.~C., {Moore}, R.~L., \& {Gary}, G.~A. 2014, \apj,
  783, 11

\bibitem[{{Alexander} \& {Fletcher}(1999)}]{1999alex}
{Alexander}, D., \& {Fletcher}, L. 1999, \solphys, 190, 167

\bibitem[{{Anzer} \& {Heinzel}(2005)}]{2005anzer}
{Anzer}, U., \& {Heinzel}, P. 2005, \apj, 622, 714

\bibitem[{{Archontis} {et~al.}(2010){Archontis}, {Tsinganos}, \&
  {Gontikakis}}]{2010archontis}
{Archontis}, V., {Tsinganos}, K., \& {Gontikakis}, C. 2010, \aap, 512, L2

\bibitem[{{Canfield} {et~al.}(1996){Canfield}, {Reardon}, {Leka}, {Shibata},
  {Yokoyama}, \& {Shimojo}}]{1996canfield}
{Canfield}, R.~C., {Reardon}, K.~P., {Leka}, K.~D., {et~al.} 1996, \apj, 464,
  1016

\bibitem[{{Carlsson} {et~al.}(2015){Carlsson}, {Leenaarts}, \& {De
  Pontieu}}]{2015carlsson}
{Carlsson}, M., {Leenaarts}, J., \& {De Pontieu}, B. 2015, \apjl, 809, L30

\bibitem[{{Chae} {et~al.}(1999){Chae}, {Qiu}, {Wang}, \& {Goode}}]{1999chae}
{Chae}, J., {Qiu}, J., {Wang}, H., \& {Goode}, P.~R. 1999, \apjl, 513, L75

\bibitem[{{Chen} {et~al.}(2015){Chen}, {Su}, {Yin}, {Priya}, {Zhang}, {Liu},
  {Xu}, \& {Yu}}]{2015chenJ}
{Chen}, J., {Su}, J., {Yin}, Z., {et~al.} 2015, \apj, 815, 71

\bibitem[{{Chen} {et~al.}(2013){Chen}, {Ip}, \& {Innes}}]{2013chen}
{Chen}, N., {Ip}, W.-H., \& {Innes}, D. 2013, \apj, 769, 96

\bibitem[{{Cheung} {et~al.}(2015){Cheung}, {De Pontieu}, {Tarbell}, {Fu},
  {Tian}, {Testa}, {Reeves}, {Mart{\'{\i}}nez-Sykora}, {Boerner}, {W{\"u}lser},
  {Lemen}, {Title}, {Hurlburt}, {Kleint}, {Kankelborg}, {Jaeggli}, {Golub},
  {McKillop}, {Saar}, {Carlsson}, \& {Hansteen}}]{2015cheung}
{Cheung}, M.~C.~M., {De Pontieu}, B., {Tarbell}, T.~D., {et~al.} 2015, \apj,
  801, 83

\bibitem[{{Chifor} {et~al.}(2008){Chifor}, {Isobe}, {Mason}, {Hannah}, {Young},
  {Del Zanna}, {Krucker}, {Ichimoto}, {Katsukawa}, \& {Yokoyama}}]{2008chifor}
{Chifor}, C., {Isobe}, H., {Mason}, H.~E., {et~al.} 2008, \aap, 491, 279

\bibitem[{{Cirtain} {et~al.}(2007){Cirtain}, {Golub}, {Lundquist}, {van
  Ballegooijen}, {Savcheva}, {Shimojo}, {DeLuca}, {Tsuneta}, {Sakao}, {Reeves},
  {Weber}, {Kano}, {Narukage}, \& {Shibasaki}}]{2007cirtain}
{Cirtain}, J.~W., {Golub}, L., {Lundquist}, L., {et~al.} 2007, Science, 318,
  1580

\bibitem[{{De Pontieu} {et~al.}(2009){De Pontieu}, {Hansteen}, {McIntosh}, \&
  {Patsourakos}}]{2009pontieu}
{De Pontieu}, B., {Hansteen}, V.~H., {McIntosh}, S.~W., \& {Patsourakos}, S.
  2009, \apj, 702, 1016

\bibitem[{{De Pontieu} {et~al.}(2014){De Pontieu}, {Title}, {Lemen}, {Kushner},
  {Akin}, {Allard}, {Berger}, {Boerner}, {Cheung}, {Chou}, {Drake}, {Duncan},
  {Freeland}, {Heyman}, {Hoffman}, {Hurlburt}, {Lindgren}, {Mathur}, {Rehse},
  {Sabolish}, {Seguin}, {Schrijver}, {Tarbell}, {W{\"u}lser}, {Wolfson},
  {Yanari}, {Mudge}, {Nguyen-Phuc}, {Timmons}, {van Bezooijen}, {Weingrod},
  {Brookner}, {Butcher}, {Dougherty}, {Eder}, {Knagenhjelm}, {Larsen},
  {Mansir}, {Phan}, {Boyle}, {Cheimets}, {DeLuca}, {Golub}, {Gates}, {Hertz},
  {McKillop}, {Park}, {Perry}, {Podgorski}, {Reeves}, {Saar}, {Testa}, {Tian},
  {Weber}, {Dunn}, {Eccles}, {Jaeggli}, {Kankelborg}, {Mashburn}, {Pust},
  {Springer}, {Carvalho}, {Kleint}, {Marmie}, {Mazmanian}, {Pereira}, {Sawyer},
  {Strong}, {Worden}, {Carlsson}, {Hansteen}, {Leenaarts}, {Wiesmann},
  {Aloise}, {Chu}, {Bush}, {Scherrer}, {Brekke}, {Martinez-Sykora}, {Lites},
  {McIntosh}, {Uitenbroek}, {Okamoto}, {Gummin}, {Auker}, {Jerram}, {Pool}, \&
  {Waltham}}]{2014IRIS}
{De Pontieu}, B., {Title}, A.~M., {Lemen}, J.~R., {et~al.} 2014, \solphys, 289,
  2733

\bibitem[{{Innes} {et~al.}(2016){Innes}, {Bucik}, {Guo}, \&
  {Nitta}}]{2016innes}
{Innes}, D., {Bucik}, R., {Guo}, L.-J., \& {Nitta}, N. 2016, ArXiv e-prints,
  arXiv:1603.03258

\bibitem[{{Innes} {et~al.}(2003){Innes}, {McKenzie}, \& {Wang}}]{2003innes}
{Innes}, D.~E., {McKenzie}, D.~E., \& {Wang}, T. 2003, \solphys, 217, 247

\bibitem[{{Jiang} {et~al.}(2007){Jiang}, {Chen}, {Li}, {Shen}, \&
  {Yang}}]{2007jiang}
{Jiang}, Y.~C., {Chen}, H.~D., {Li}, K.~J., {Shen}, Y.~D., \& {Yang}, L.~H.
  2007, \aap, 469, 331

\bibitem[{{Leenaarts} {et~al.}(2013{\natexlab{a}}){Leenaarts}, {Pereira},
  {Carlsson}, {Uitenbroek}, \& {De Pontieu}}]{2013a_leenaarts}
{Leenaarts}, J., {Pereira}, T.~M.~D., {Carlsson}, M., {Uitenbroek}, H., \& {De
  Pontieu}, B. 2013{\natexlab{a}}, \apj, 772, 89

\bibitem[{{Leenaarts} {et~al.}(2013{\natexlab{b}}){Leenaarts}, {Pereira},
  {Carlsson}, {Uitenbroek}, \& {De Pontieu}}]{leenaarts_2013b}
---. 2013{\natexlab{b}}, \apj, 772, 90

\bibitem[{{Lemen} {et~al.}(2012){Lemen}, {Title}, {Akin}, {Boerner}, {Chou},
  {Drake}, {Duncan}, {Edwards}, {Friedlaender}, {Heyman}, {Hurlburt}, {Katz},
  {Kushner}, {Levay}, {Lindgren}, {Mathur}, {McFeaters}, {Mitchell}, {Rehse},
  {Schrijver}, {Springer}, {Stern}, {Tarbell}, {Wuelser}, {Wolfson}, {Yanari},
  {Bookbinder}, {Cheimets}, {Caldwell}, {Deluca}, {Gates}, {Golub}, {Park},
  {Podgorski}, {Bush}, {Scherrer}, {Gummin}, {Smith}, {Auker}, {Jerram},
  {Pool}, {Soufli}, {Windt}, {Beardsley}, {Clapp}, {Lang}, \&
  {Waltham}}]{2012lemen}
{Lemen}, J.~R., {Title}, A.~M., {Akin}, D.~J., {et~al.} 2012, \solphys, 275, 17

\bibitem[{{Liu} {et~al.}(2015){Liu}, {Heinzel}, {Kleint}, \& {Ka{\v
  s}parov{\'a}}}]{2015liu}
{Liu}, W., {Heinzel}, P., {Kleint}, L., \& {Ka{\v s}parov{\'a}}, J. 2015,
  \solphys, 290, 3525

\bibitem[{{Moore} {et~al.}(2010){Moore}, {Cirtain}, {Sterling}, \&
  {Falconer}}]{2010Moore}
{Moore}, R.~L., {Cirtain}, J.~W., {Sterling}, A.~C., \& {Falconer}, D.~A. 2010,
  \apj, 720, 757

\bibitem[{{Mulay} {et~al.}(2016){Mulay}, {Tripathi}, {Del Zanna}, \&
  {Mason}}]{2016mulay}
{Mulay}, S.~M., {Tripathi}, D., {Del Zanna}, G., \& {Mason}, H. 2016, \aap,
  589, A79

\bibitem[{{Newton} {et~al.}(1995){Newton}, {Emslie}, \& {Mariska}}]{1995newton}
{Newton}, E.~K., {Emslie}, A.~G., \& {Mariska}, J.~T. 1995, \apj, 447, 915

\bibitem[{{Nishizuka} {et~al.}(2008){Nishizuka}, {Shimizu}, {Nakamura},
  {Otsuji}, {Okamoto}, {Katsukawa}, \& {Shibata}}]{2008nishizuka}
{Nishizuka}, N., {Shimizu}, M., {Nakamura}, T., {et~al.} 2008, \apjl, 683, L83

\bibitem[{{Pesnell} {et~al.}(2012){Pesnell}, {Thompson}, \&
  {Chamberlin}}]{pesnell2015solar}
{Pesnell}, W.~D., {Thompson}, B.~J., \& {Chamberlin}, P.~C. 2012, \solphys,
  275, 3

\bibitem[{{Schmieder} {et~al.}(1994){Schmieder}, {Golub}, \&
  {Antiochos}}]{1994schmieder}
{Schmieder}, B., {Golub}, L., \& {Antiochos}, S.~K. 1994, \apj, 425, 326

\bibitem[{{Schou} {et~al.}(2012){Schou}, {Scherrer}, {Bush}, {Wachter},
  {Couvidat}, {Rabello-Soares}, {Bogart}, {Hoeksema}, {Liu}, {Duvall}, {Akin},
  {Allard}, {Miles}, {Rairden}, {Shine}, {Tarbell}, {Title}, {Wolfson},
  {Elmore}, {Norton}, \& {Tomczyk}}]{2012schou}
{Schou}, J., {Scherrer}, P.~H., {Bush}, R.~I., {et~al.} 2012, \solphys, 275,
  229

\bibitem[{{Shibata} {et~al.}(1992){Shibata}, {Ishido}, {Acton}, {Strong},
  {Hirayama}, {Uchida}, {McAllister}, {Matsumoto}, {Tsuneta}, {Shimizu},
  {Hara}, {Sakurai}, {Ichimoto}, {Nishino}, \& {Ogawara}}]{1992shibata}
{Shibata}, K., {Ishido}, Y., {Acton}, L.~W., {et~al.} 1992, \pasj, 44, L173

\bibitem[{{Shibata} {et~al.}(2007){Shibata}, {Nakamura}, {Matsumoto}, {Otsuji},
  {Okamoto}, {Nishizuka}, {Kawate}, {Watanabe}, {Nagata}, {UeNo}, {Kitai},
  {Nozawa}, {Tsuneta}, {Suematsu}, {Ichimoto}, {Shimizu}, {Katsukawa},
  {Tarbell}, {Berger}, {Lites}, {Shine}, \& {Title}}]{2007shibata}
{Shibata}, K., {Nakamura}, T., {Matsumoto}, T., {et~al.} 2007, Science, 318,
  1591

\bibitem[{{Shimojo} {et~al.}(1996){Shimojo}, {Hashimoto}, {Shibata},
  {Hirayama}, {Hudson}, \& {Acton}}]{1996Shimojo}
{Shimojo}, M., {Hashimoto}, S., {Shibata}, K., {et~al.} 1996, \pasj, 48, 123

\bibitem[{{Sterling} {et~al.}(2015){Sterling}, {Moore}, {Falconer}, \&
  {Adams}}]{2015sterling}
{Sterling}, A.~C., {Moore}, R.~L., {Falconer}, D.~A., \& {Adams}, M. 2015,
  \nat, 523, 437

\bibitem[{{Winebarger} {et~al.}(1999){Winebarger}, {Emslie}, {Mariska}, \&
  {Warren}}]{1999winebarger}
{Winebarger}, A.~R., {Emslie}, A.~G., {Mariska}, J.~T., \& {Warren}, H.~P.
  1999, \apj, 526, 471

\bibitem[{{Winebarger} {et~al.}(2002){Winebarger}, {Emslie}, {Mariska}, \&
  {Warren}}]{2002winebarger}
---. 2002, \apj, 565, 1298

\bibitem[{{Yokoyama} \& {Shibata}(1995)}]{1995yokoyama}
{Yokoyama}, T., \& {Shibata}, K. 1995, \nat, 375, 42

\bibitem[{{Zeng} {et~al.}(2016){Zeng}, {Chen}, {Ji}, {Goode}, \&
  {Cao}}]{2016zeng}
{Zeng}, Z., {Chen}, B., {Ji}, H., {Goode}, P.~R., \& {Cao}, W. 2016, \apjl,
  819, L3

\bibitem[{{Zhang} \& {Ji}(2014)}]{2014zhang}
{Zhang}, Q.~M., \& {Ji}, H.~S. 2014, \aap, 567, A11

\end{thebibliography}
\end{document}